# Physics with Like-Sign Muon Beams in a TeV Muon Collider

Clemens A. Heusch* and Frank Cuypers†

*Santa Cruz Institute for Particle Physics
University of California, Santa Cruz
†Max-Planck-Institut für Physik
Föhringer Ring 6, D-80805 Munich, Germany

**Abstract.** We point out that both the specific lepton number content and the high energies potentially attainable with muon-muon colliders make it advisable to consider the technical feasibility of including an option of like-sign incoming beams in the studies towards a proposal to build a muon-muon collider with center-of-mass energies in the TeV region. This capability will add some unique physics capabilities to the project. Special attention will have to be paid to polarization retention for the muons.

## INTRODUCTION

The prospect of having lepton colliders reach into the TeV range has opened up a slew of fascinating physics perspectives: in a number of workshops through the past several years, electron linear colliders have been investigated in terms of the physics programs that motivate a choice of parameters—such as center-of-mass energies, luminosities, polarization parameters, and the precise particle choice for the initial state - $e^+e^-$, $e^-e^-$, $e\gamma$, $\gamma\gamma$. For a number of investigations, particularly those that probe the limits of the Standard Model and reach beyond its boundaries, it has become increasingly clear that it is vitally important to go to the highest attainable energies. Also, a clear definition of incoming polarization states defines the process to be studied and suppresses backgrounds, and is needed for many if not most studies of subtle points in new interactions.

In the framework of machine studies, electron colliders beyond LEP-2 energies are relegated to a linear configuration, thus putting a fiscal limit to the energies that can reasonably be reached (quite apart from the beamstrahlung effects that will exacerbate experimentation at the upper end of the attainable energies). Muon-muon colliders, however, are being proposed well beyond the 2-TeV limit that is practically imposed by economic considerations. This means that interactions that come into their own only well above 1 TeV, and which do not depend on the specific lepton family tag of the incoming beams, are more promising in a muon collider, while some new processes will profit from being studied both in an electron–electron and in a muon–muon system. In this presentation, I will show that both arguments apply for some of the most exciting physics spoils we might expect from the electron-electron mode of a projected NLC.

## A LIKE-SIGN MUON COLLIDER

It has been shown [1,2] that, in the electron collider case, an $e^-e^-$ option is a natural ingredient in a broad-based facility for experimentation: It poses no accelerator problems with the possible exception of beam-beam disruption at the interaction point. It has the added advantage that its beams can both be highly polarized - a feature not clearly realizable in the $e^+e^-$ option - certainly not with trivial polarization reversibility. In addition, the $e$–$\gamma$ and $\gamma$–$\gamma$ options that are being planned at electron colliders, depend heavily on two highly polarized incoming lepton beams for both spectrum and polarization of the photons. What are the corresponding virtues and difficulties of installing a like-sign option ($\mu^+\mu^+$ or $\mu^-\mu^-$) in the muon collider?

For the argument's sake, I will take D. Neuffer's footprint for a muon collider facility [3] to point out the few essential features that have to be kept in mind for an overall evaluation of the feasibility, at affordable cost, of a like-sign option. Take, e.g., the variant where a muon injector linac feeds the phase-space-compressed muons into a rapid-cycling synchrotron as the principal accelerator: here, like-sign operation would simplify everything except the final collider ring, in which a second beam pipe and guidance system have to be installed. This is not cheap, but presents no technical problem.

We see the main technical challenge in the physics need for highly polarized muon beams—just like the high degree of polarization today available in electron linacs: Muons originate in weak decay with a well-defined helicity—but in the obvious attempt to maximize the muon flux, forward- and backward-emitted muons will be mixed in the cooling and compression process: downstream, the largest fluxes will be a mixture of two helicities. It should not take a great deal of ingenuity, however, to narrow the acceptance of the cooling devices such as to admit only one helicity state, at the expense of a flux loss well below a factor of two. This is a development project that imposes itself, as we will see below from the stress on the optimal definition of helicity states for cogent physics studies of, above all, Beyond-the-Standard-Model processes. It will handsomely pay off in a fuller definition of the chiral couplings that permit or enhance the processes we are trying to study, and thereby have a decisive influence on signal definition and background suppression. These aspects have been fully investigated in the framework of electron–electron scattering, where the additional feature of easy reversal or switching off of a high degree of polarization permits a convincing check on promising but statistically unconvincing signals. Like-sign muon colliders have the advantage of higher energy reach, but will have to strive hard for the added thrust of polarization definition.

In the following, we pass review of a few processes that are sure to benefit greatly from the energy reach of the muon collider as well as from the "exotic" charge, weak isospin, and lepton flavor and number of the incoming channel. While largely based on work performed in more quantitative fashion for the electron-electron version of an NLC, detailed calculations and modelling are straightforward.

## STRONG GAUGE BOSON INTERACTIONS

Should the Standard Model go unchallenged by new-physics signals up to the TeV level without producing evidence for an elementary Higgs boson, we expect a new regime of the strong interaction to manifest itself [4]. The attendant new term in the effective Lagrangian, frequently denoted $L_5$, breaks local gauge symmetry, and becomes observable as the longitudinal components of the $W$ and $Z$ bosons act as Goldstone scalars. An experimental investigation of their interactions then becomes as fundamentally important as that of the pions in the 100 MeV region, and has to be pursued in all J and I channels. $e^+e^-$ annihilation gives evidence only in the J=1, I=0,1 channels; gamma gamma scattering adds data on the J=0,2 channels. The potentially distinctive I=2 channel is accessible through quark–antiquark or gluon fusion in hadron colliders, but with massive attendant backgrounds from heavy-quark decays. While strong $WW$ scattering may not produce easily identifiable signals at $e^+e^-$ colliders, like-sign $ee$ or $\mu\mu$ collisions could unveil a new regime of dynamics that manifests itself, e.g., by the formation of vector $\rho$-type resonances [5]. Figure 1 shows the basic diagram, where only unobserved escaping neutrinos limit the full reconstruction of the final state. While $e^-e^-$ colliders have the distinctive feature that we can switch polarization parameters essentially at will, like-sign muon colliders with their superior energy reach may well be needed for a full exploration of these new phenomena.

In addition to these dynamical ideas, an extended Higgs sector favors among possible models one version with fundamental doublets and triplets [6] that can lead to easily recognized sharp structure in the s-channel mass plot of final-state W-pair distributions that are emitted isotropically in their center-

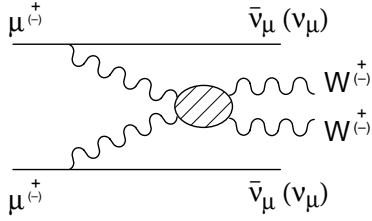

**FIGURE 1.** Strong $WW$ scattering has a clean signature in $\mu^{+(-)}\mu^{+(-)}$ scattering: only the missing transverse momentum of the escaping neutrinos is absent in the $WW$ final-state observation.

of-mass. This evidence, due to the process

$$\begin{aligned}\mu^{+(-)}\mu^{+(-)} &\to \overset{(-)}{\nu_\mu}\overset{(-)}{\nu_\mu} W^{+(-)} W^{+(-)} \\ &\to \overset{(-)}{\nu_\mu}\overset{(-)}{\nu_\mu} H^{++(--)}\end{aligned} \quad (1)$$

where the $WW$ fusion process leads to $H^{++(--)}$ formation with a full unit of R in cross-section (Fig. 2b), could be quite spectacular, as shown in Fig. 3. What increases the interest in performing this search at a muon collider is the possibility that a direct coupling of two like-sign leptons to the $H^{++(--)}$, shown in Fig. 2a, of a theoretically unpredictable strength, might well differ between $ee$ and $\mu\mu$ collisions. This feature adds spice to the search in the $\mu\mu$ case, even if there were prior evidence for such a discovery in $e^-e^-$ experimentation.

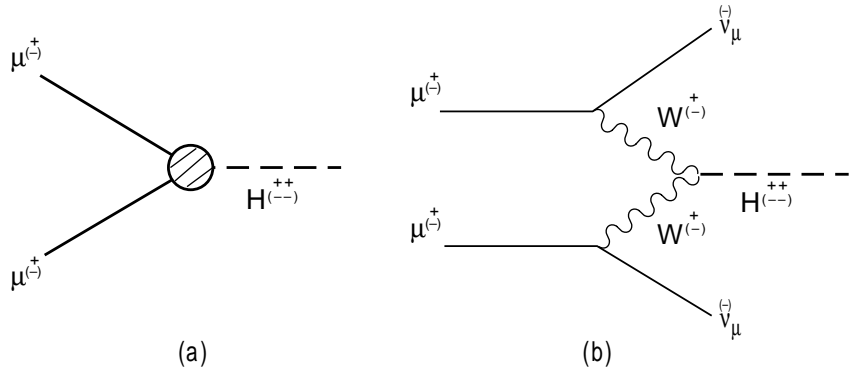

**FIGURE 2.** For the production of a doubly charged Higgs boson, the $W^{+(-)}W^{+(-)}$ fusion (graph b) will yield a full unit of $R[(=\sigma(e^+e^-\to\mu^+\mu^-)]$ in cross section. The unknown coupling $\mu\mu H$ in graph (a) may or may not differ from a direct $eeH$ coupling, if it is not hopelessly suppressed.

**FIGURE 3.** Invariant mass distribution for $W^{+(-)}W^{+(-)}$ scattering in the presence of a doubly charged Higgs meson of mass 0.2 or 0.3 TeV (from Ref. [6]).

# NEW CONTACT INTERACTIONS

When we enter into a new energy regime of a particle interaction, one question that imposes itself is: Does a simple interaction, like muon-muon Møller scattering, show any signs that the simple QED Lagrangian has to add a new term characterized by the exchange of a heavy gauge boson with a mass well above what would have manifested itself before? Buchmüller and Wyler [7] showed that such a new interaction, with $\Lambda \gg \sqrt{s}$, can be accommodated by a minimal contact term in the Lagrangian

$$\mathcal{L}_{\text{eff}} \sim \frac{g^2}{\Lambda^2} \eta_{LL} \left(\overline{\psi}_L^1 \gamma_\mu \psi_L^1\right) \left(\overline{\psi}_L^2 \gamma^\mu \psi_L^2\right) + \eta_{RR} \left(\overline{\psi}_R^1 \gamma_\mu \psi_R^1\right) \left(\overline{\psi}_R^2 \gamma^\mu \psi_R^2\right) \\ + \eta_{RL} \left(\overline{\psi}_R^1 \gamma_\mu \psi_R^1\right) \left(\overline{\psi}_L^2 \gamma^\mu \psi_L^2\right) + \eta_{LR} \left(\overline{\psi}_L^1 \gamma_\mu \psi_L^1\right) \left(\overline{\psi}_R^2 \gamma^\mu \psi_R^2\right), \quad (2)$$

with $\psi_R, \psi_L$ the standard chirality projections of the electron spinors, $\psi_L = \frac{1}{2}(1 - \gamma_5)\psi$, $\psi_R = \frac{1}{2}(1 + \gamma_5)\psi$. If we want to test the point-like character of the muon, we can study the appropriate sensitivity of the Møller scattering cross section to a compositeness scale $\Lambda$ as demonstrated in Fig. 4 [8]. Note that interference of crossing terms makes the Møller cross section more sensitive than the "Bhabha"-type $\mu^+\mu^-$ case, even in the absence of polarization (Fig. 4a). At higher energy, and for high degrees of polarizations, the like-sign scattering sensitivity is able to reach impressive new-interaction energy scales for luminosities that are scaled to those reached in the PETRA investigations of present compositeness limits.

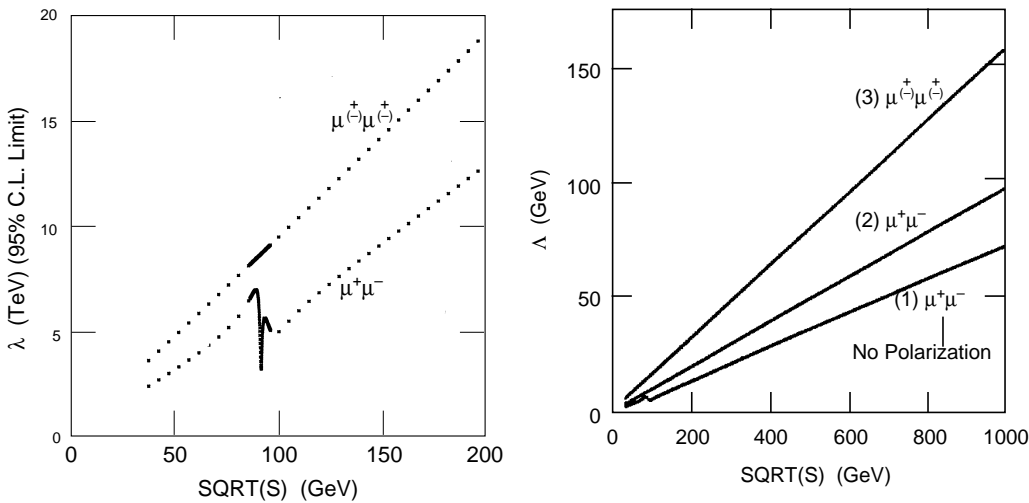

**FIGURE 4.** a) The sensitivity of Møller vs. Bhabha scattering below and above the $Z$ resonance, in the absence of polarization of the incoming beams, to a compositeness scale $\Lambda$ [TeV]. b) The corresponding sensitivity comparison in the $\mu^+\mu^-$ case where none (curve 1) or one beam (curve 2) are polarized, and like-sign scattering with both beams polarized (curve 3). The luminosity is chosen such that the statistics at each energy corresponds to the integrated values reached at PETRA for the compositeness limits set there.

Analogous investigations can be performed on the influence of other massive new boson exchanges, like heavy $Z'$ states [9].

# HEAVY MAJORANA NEUTRINOS

It has long been pointed out [10–13] that like-sign electron scattering can produce quasi-elastic $W$ pairs, either real or virtual, via Majorana neutrino exchange. While this lepton-number and -flavor violating process (Fig. 5) has been treated mainly in the Left–Right-Symmetric Model, Heusch and Minkowski recently showed [14] that a minimal neutrino mass generating case can be formulated that permits sizeable cross-sections for left-handed $WW$ production in quasi-elastic $e^-e^- \to W^-W^-$ reactions according to Fig. 5: the couplings of the incoming leptons to the exchanged heavy neutrino(s) contain specific mass mixing matrices: clearly, there is a chance that the case where incoming muons have to couple to the exchanged TeV-level Majorana neutrino may be dominated by matrix elements $\cup_{\mu\alpha}$ different from those prevailing in the $e^-e^-$ case.

**FIGURE 5.** Lowest-order graphs for the process $\mu^{(-)}_{+}\mu^{(-)}_{+} \to W^{(-)}_{+}W^{(-)}_{+}$, mediated by the exchange of a heavy Majorana neutrino.

While this fact in itself is of interest in terms of a novel discovery chance, we stress that the cross section for the process

$$\mu^{(-)}_{+}\mu^{(-)}_{+} \xrightarrow{N_\alpha} W^{(-)}_{+} W^{(-)}_{+} \qquad (3)$$

has a hard-scattering term proportional to $s^2$,

$$\sigma^{(N)} = \frac{G_F}{16\pi} \frac{s^2}{m^2_{red}} |\eta_N|^2,$$

$$\text{with } \frac{1}{m_{red}} = \sum_\alpha \frac{1}{m_\alpha} \quad (\alpha \text{ is the high} - \text{mass neutrino index}) \qquad (4)$$

$$\text{and } \eta_N = \sum_\alpha (\cup_{\mu\alpha})^2 \frac{m_{red}}{m_\alpha}.$$

As long as $\sqrt{s} < m(N)$, with $N$ the exchanged heavy neutrino, this cross section will enormously profit from the increased energy range provided by the muon collider when compared with its electron-accelerating analog.

We have not yet explored the details of the change in the theoretical treatment of the active mass matrix elements for the present case. But we emphasize that while electron-initiated reactions (2) are, at least in principle, also present in neutrinoless double beta decay [15], no such equivalence exists in the muon-muon collision case. We therefore believe it to be a valuable discovery aim to look for the potentially spectacular and unmistakable final-state configurations (back-to-back $W^-$ pairs; unlike-flavor, almost back-to-back-emitted negative leptons with missing momenta from escaping light neutrinos, etc. in the otherwise crowded final states of violent muon-muon collisions.

## SUPERSYMMETRY

We now turn to the framework of the minimal supersymmetric Standard Model [16]. This implies making use of two *essential* assumptions: (i) *R*- parity invariance;[1] (ii) the lightest supersymmetric particle is a neutralino, $\tilde{\chi}_1^0$, it is *stable* and interacts *weakly* with matter, *i.e.*, it escapes detection. If these conditions are not fulfilled, the results of the analysis to follow are qualitatively incorrect.

The pair-production of like-sign smuons in $\mu^-\mu^-$ collisions

$$\mu^-\mu^- \to \tilde{\mu}^-\tilde{\mu}^- \qquad (5)$$

and their decays into muons and invisible particles

$$\tilde{\mu}^- \to \mu^- \; \tilde{\chi}_1^0 \qquad (6)$$

$$\begin{aligned}\tilde{\mu}^- \to \mu^- \; &\tilde{\chi}_2^0 \\ &\hookrightarrow \tilde{\chi}_1^0 \; Z^0 \\ &\qquad \hookrightarrow \nu\bar{\nu}\end{aligned} \qquad (7)$$

$$\begin{aligned}\tilde{\mu}^- \to \nu \; &\tilde{\chi}_1^- \\ &\hookrightarrow \tilde{\chi}_1^0 \; W^- \\ &\qquad \hookrightarrow \bar{\nu}\mu^-\end{aligned} \qquad (8)$$

$$\vdots$$

are depicted in the Feynman diagram of Fig. 6. They proceed exactly in the same way as for selectrons in $e^-e^-$ collisions. The whole analysis of Refs. [17,18] can thus be applied here, with higher energies and without polarization.

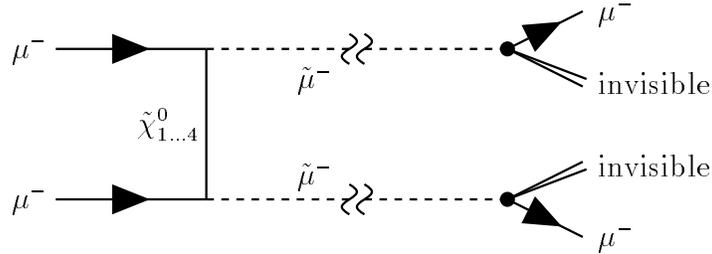

**FIGURE 6**. Feynman diagram for the production and decay of smuons.

In Fig. 7 we show the energy behavior of the smuon pair production cross section [17] times the branching fractions of these smuons into muons and invisible particles [18] for different smuon masses. For definiteness we have chosen the other supersymmetry parameters to take some "typical" values

$$\tan\beta = 4 \qquad \mu = 500 \text{ GeV} \qquad M_2 = 500 \text{ GeV} \qquad (9)$$

---

[1]If this *ad hoc* symmetry is broken, the supersymmetric phenomenology becomes actually much easier because lepton number violating processes should then show up.

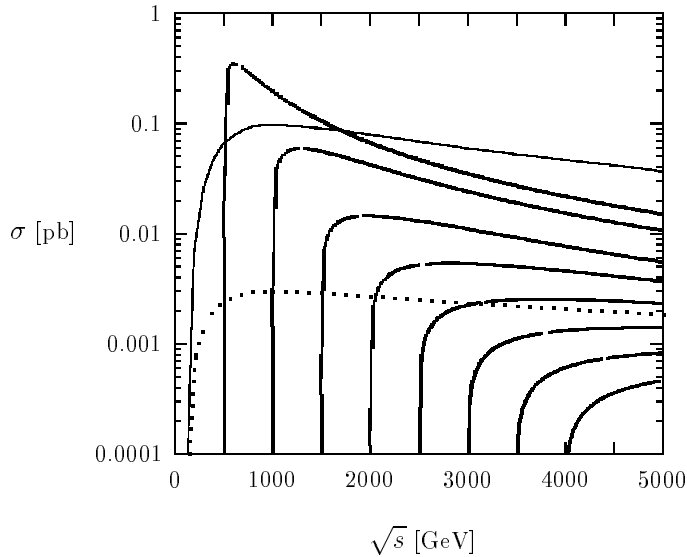

**FIGURE 7.** Smuon pair-production and decay cross section as a function of the collider energy for smuon masses ranging from 250 GeV to 2000 GeV. The Standard Model background is shown by the thinner line. The dotted line represents the signal cross section needed to exceed three times the Poisson error of the background with 100 fb$^{-1}$ of data.

and have assumed the soft supersymmetry breaking terms to have a common value at the grand unified scale. The corresponding mass of the lightest neutralino is 244 GeV; it is predominantly a mixture of photino and zino. The other neutralinos and charginos have masses around 500 GeV. The asymptotic behavior of the cross section is

$$\sigma \sim \frac{\pi\alpha^2}{2s} \ln \frac{s}{\Lambda^2_{\text{SUSY}}} \ . \qquad (10)$$

The expected Standard Model background, mainly from $W^-$ bremsstrahlung, is also shown in Fig. 7. To evaluate this background, we have imposed the mild detector acceptance cuts $|\eta_\mu| < 2$ and $E_\mu > \sqrt{s}/10$ on the emerging muons. Close to threshold, *i.e.*, for maximum signal, these cuts do not affect the smuon cross sections.

Operating the collider at 1 TeV, smuons up to almost 500 GeV can be produced with a signal-to-noise ratio of one. Insisting that the supersymmetric signal exceeds the Standard Model background by at least three standard deviations, we find, with a luminosity of 100 fb$^{-1}$ and a collider energy of 3 TeV, that smuons up to 1250 GeV can be discovered just by comparing total rates. For heavier smuons, a more subtle analysis will be required. It must be noted, however, that such heavy smuons are not favored at all by low energy supersymmetry.

# ANOMALOUS QUARTIC GAUGE COUPLINGS

The non-abelian part of the electroweak sector remains up to now a little-explored area of the Standard Model. Although existing LEP I data are already testing the trilinear Yang-Mills couplings through loops [19], no effects of the quartic gauge interactions have yet been observed.

New physics at the TeV scale is bound to induce some anomalies in the gauge sector. The exact form of these deviations from the Standard Model expectations depends of course on the precise nature of unknown phenomena, and it is customary to parametrize our ignorance in terms of effective lagrangians. There are infinitely many operators which can induce non-standard vector boson couplings. However, those which have the lowest dimension and which do not induce any trilinear couplings are expected to yield the largest contributions, because they can result from tree-level exchanges of heavy particles. There are only two such so-called "genuine" quartic operators of dimension four. They modify the $W^4$ and $W^2Z^2$ vertices and induce a novel $Z^4$ vertex, by adding the following pieces to the Standard Model lagrangian [20]:

$$\mathcal{L}_0 = g_0 g_W^2 \left( W^{+\mu} W_\mu^- W^{+\nu} W_\nu^- \right.$$

$$\left. + \frac{1}{\cos^2 \theta_w} W^{+\mu} W_\mu^- Z^\nu Z_\nu + \frac{1}{4\cos^4 \theta_w} Z^\mu Z_\mu Z^\nu Z_\nu \right) ,$$

(11)

$$\mathcal{L}_c = g_c g_W^2 \left[ \frac{1}{2}(W^{+\mu} W_\mu^- W^{+\nu} W_\nu^- + W^{+\mu} W_\mu^+ W^{-\nu} W_\nu^-) \right.$$

$$\left. + \frac{1}{\cos^2 \theta_w} Z^\mu W_\mu^+ Z^\nu W_\nu^- + \frac{1}{4\cos^4 \theta_w} Z^\mu Z_\mu Z^\nu Z_\nu \right] ,$$

where $\theta_w$ is the weak mixing angle, $g_W$ is the usual $W$ coupling and $g_0$ and $g_c$ parametrize the strength of the anomalies. Note that no anomalous quartic operators of dimension four involve the photon.

As in $e^- e^-$ scattering [21], such anomalous couplings would show up dramatically in high energy $\mu^- \mu^-$ reactions such as

$$\mu^- \mu^- \to \nu_\mu \nu_\mu W^- W^- , \tag{12}$$

$$\mu^- \mu^- \to \mu^- \mu^- W^+ W^- , \tag{13}$$

$$\mu^- \mu^- \to \mu^- \nu_\mu Z^0 W^- , \tag{14}$$

$$\mu^- \mu^- \to \mu^- \mu^- Z^0 Z^0 , \tag{15}$$

because they spoil the normally very effective unitarity cancelations between the diagrams involving the quartic vertices and the diagrams. Even tiny values of $g_0$ or $g_c$ provoke significant increases of the total cross sections, the more so at higher energies.

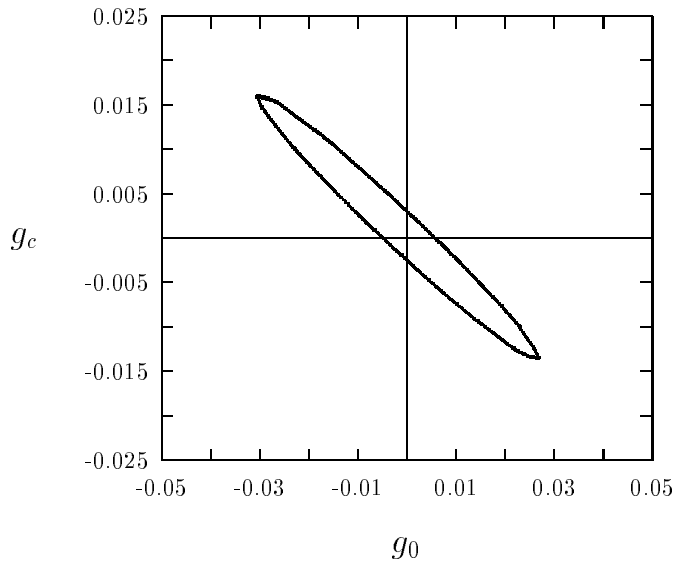

**FIGURE 8.** Contours of observability at 95% confidence level of the anomalous quartic gauge couplings $g_0$ and $g_c$, for the reaction (11) at 2 TeV and with 100 fb$^{-1}$ of data.

Assuming 100 fb$^{-1}$ of accumulated luminosity at a 2 TeV $\mu^-\mu^-$ facility, and concentrating solely on the hadronic decays of the $W$'s, we show in Fig. 8 the area in the $(g_0, g_c)$ plane beyond which these anomalies would be observed in reaction (11) with better than 95% confidence. The strong correlation can be lifted by also studying reactions (12-14) [21]. The anomalous parameters should then be constrainable down to a few tenths of a percent.

## CONCLUSION

In the preceding brief discussion, we have shown how muon-muon scattering in the TeV region can be considerably enriched by making like-sign beams available as the input channel. The list we give is indicative rather than complete. It becomes evident that a major effort towards making these beams well-defined in their helicity states is easily motivated by the physics promise, and we urge an early feasibility study for implementation of a like-sign version, including a well-defined high degree of polarization.

## ACKNOWLEDGMENTS

We thank David Cline for encouraging this contribution to the $\mu\mu$ Workshop, and the Workshop participants for many instructive discussions about muon colliders. It is a pleasure for F.C. to thank Karol Kołodziej and Geert Jan van Oldenborgh for many fruitful discussions and their collaboration on the topics of the last two sections.